\documentclass[twocolumn,nofootinbib,prl,floatfix,preprintnumbers,amsmath,amssymb,groupedaddress,superscriptaddress]{revtex4}
\usepackage{dsfont}
\usepackage{graphicx} 
\usepackage{epstopdf} 
\usepackage{slashed}
\usepackage[bf,tight]{subfigure}

\newcommand{\GeV}{\,\mathrm{GeV}}

\newcommand{\genericT}{{\ensuremath{\rm T}}}
\newcommand{\pt}{p_{\rm T}}

\newcommand{\mpt}{\slashed{p}_\genericT}
\newcommand{\mptvec}{\slashed{\vec{p}}_\genericT}

\newcommand{\half}{{\frac{1}{2}}}

\newcommand{\definmath}[2] {\def#1{\ifmmode#2\else$#2$\fi}}
\definmath{\invfb}{\mathrm{fb}^{-1}}

\newcommand{\paper}{letter}
\newcommand{\newvar}{m_{\tau\tau}^{\mathrm{Higgs-bound}}}

\newcommand{\mtautaueff}{{m_{\tau\tau}^{\mathrm{Effective}}}}
\newcommand{\mtautauvis}{{m_{\tau\tau}^{\mathrm{Visible}}}}
\newcommand{\mttrue}{{m_{\rm T}^{\rm True}}}

\newcommand{\hideIt}[1]{{}}
\newcommand{\mysubsection}[1]{\vspace{-5mm}\subsection{#1}\vspace{-2mm}}

\begin{document}   
\title{Speedy Higgs boson discovery in decays to tau lepton pairs : $h\rightarrow\tau\tau$}
\date{\today}

\author{Alan J. Barr}
\email{a.barr@physics.ox.ac.uk}
\affiliation{Denys Wilkinson Building, Keble Road, Oxford, OX1 3RH, United Kingdom}

\author{Sky T. French}
\email{sfrench@hep.phy.cam.ac.uk}  

\author{James A. Frost}
\email{frost@hep.phy.cam.ac.uk}  

\author{Christopher G. Lester}
\email{lester@hep.phy.cam.ac.uk}
\affiliation{Cavendish Laboratory, Dept of Physics, JJ Thomson Avenue,
Cambridge, CB3 0HE, United Kingdom}

\preprint{Cavendish-HEP-11/12}
\begin{abstract} 
Discovery of the Higgs boson in any decay channel depends on the existence of event variables or cuts with sensitivity to the 
presence of the Higgs.  
We demonstrate the non-optimality of the kinematic variables which are currently expected to play the largest role in the 
discovery (or exclusion) of the Higgs at the LHC in the $\tau\tau$ channel.  Any LHC collaboration looking for opportunities to gain 
advantages over its rivals should, perhaps, consider the alternative strategy we propose.
\end{abstract}   
\maketitle 


\mysubsection{Introduction}

There is much to be gained from constructing event variables which place maximal lower-bounds on well defined quantities of interest.  Such variables can be used to select events containing new-physics when the scale of the property which is being ``bounded'' is higher in the signal than in the most important backgrounds.
\hideIt{Examples of such variables include $m_{T2}$, (which is the maximal lower-bound for the mass of (either member of) a pair of identical-mass particles in an event each having invisible daughters \cite{Lester:1999tx,Barr:2003rg,Cheng:2008hk}) or $\mttrue$ (which is the maximal lower-bound for the mass of a Higgs boson decaying to $h\rightarrow WW\rightarrow l\nu l\nu$ \cite{Barger:1987re,Barr:2009mx}) to name just two of many.} 
One may construct {\em the single} variable that bounds an arbitrary scale by considering that scale (often a mass) to be a function of all the unknowns in the event (often components of invisible particle momenta).  Having done this, the minimal value of this scale over all possible values of those unknowns, subject to any constraints that need to be asserted to enforce consistency, is the bound in question.\footnote{Examples in the context of Higgs boson searches include Refs.~\cite{Barr:2009mx,Gross:2009wia}. In Ref.~\cite{Barr:2011xt} a recent attempt has been made to formally write down the steps that are needed to construct such maximal lower-bounding variables for a wide class of circumstances.}  The transverse mass is an example of such a maximal lower-bound variable: when applied to a $W\rightarrow l\nu$ event in a hadron collider it returns {\em the} largest possible lower-bound on the $W$-mass that may be derived from that event (given access to the lepton four-momentum and the missing transverse two-momentum only) assuming that there were no confounding sources of missing transverse momentum.

\mysubsection{Separating $Z\rightarrow\tau\tau$ from $h\rightarrow\tau\tau$}

\begin{figure}
 \begin{center}
  \includegraphics[width=0.9\columnwidth,height=0.6\columnwidth]{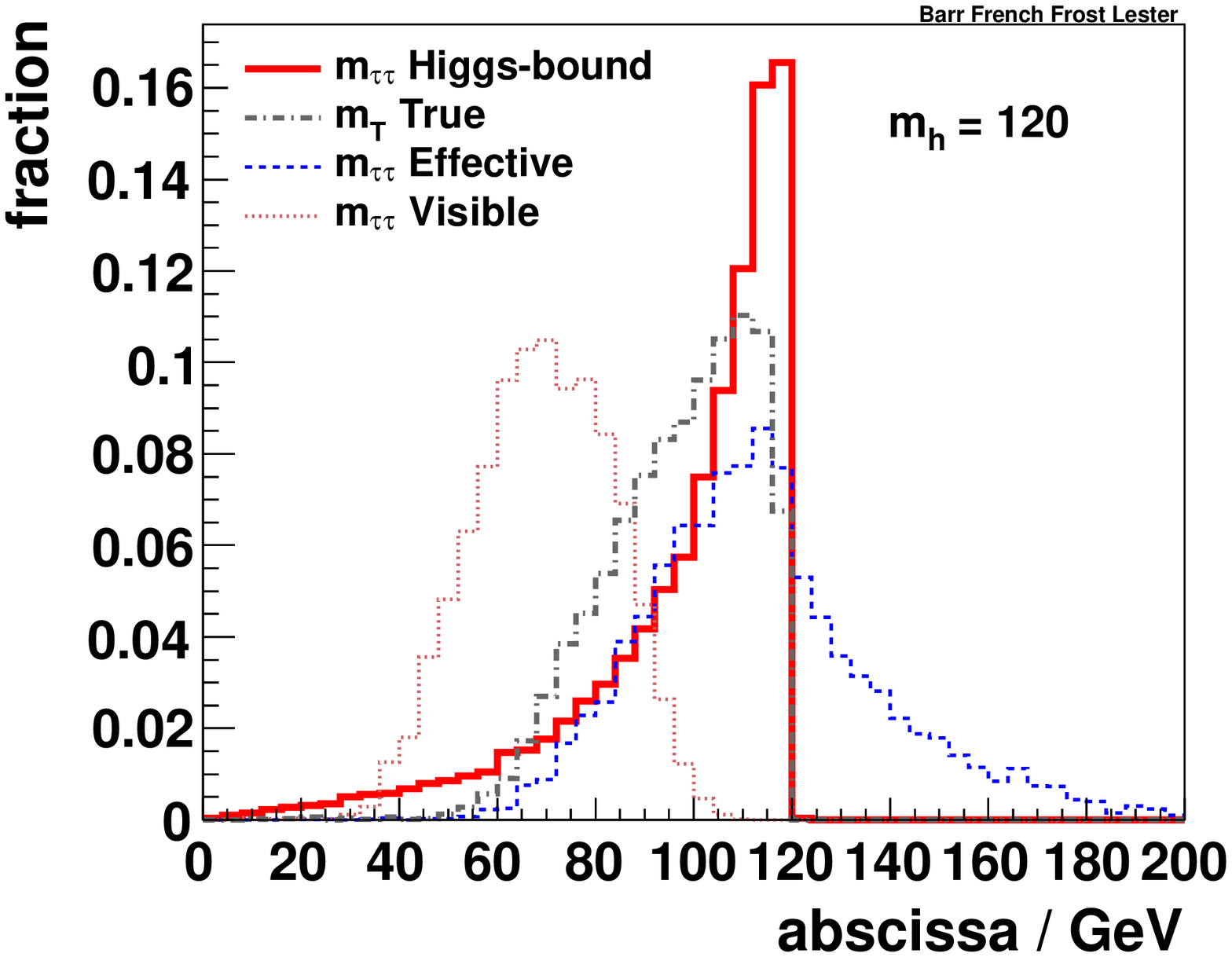}
\caption{
At Monte Carlo truth level, and for signal only, the distribution of the new $\newvar$ variable (black) is compared to the existing variables $\mttrue$ \cite{Barr:2009mx} (magenta), $\mtautaueff$ \cite{kuhn-blois} (blue) and $\mtautauvis$ \cite{ATLAS-CONF-2011-024,cms_higgs_tau_tau} (cyan) for a 120~GeV Higgs decaying to $\tau\tau$.
All histograms are scaled to unit area. 
}
\label{fig:truth}
\end{center}
\end{figure}

The main background to searches for $h\rightarrow\tau\tau$ is $Z\rightarrow\tau\tau$, so
following the general procedure described above, one would expect that the best way to separate the signal from this irreducible background is to 
construct the variable which provides the maximal lower bound for the ``parent'' mass (i.e.~$m_H$ or $m_Z$ in signal and 
background respectively) given the observed visible decay products of the taus together with the net missing transverse momentum.  
In a ``perfect'' detector, such a variable (we will call it $\newvar$) should, by construction, place all the irreducible 
background $Z\rightarrow\tau\tau$ events at values of $\newvar$ below $m_Z$, leaving the region $m_Z < \newvar\le m_h$ available to 
the signal and entirely free of background.   Following the general procedure described above, we are therefore naturally led to 
construct $\newvar$ defined as follows\footnote{The notation used in this \paper{} follows that of \cite{Barr:2011xt} exactly -- 
see in particular Tables I, II, VI and VII therein for reference.  In the specific context of the decay $h\rightarrow \tau_1 
\tau_2$ we denote the measured four momentum of the visible decay products of the harder and softer tau by $P_1^\mu$ and $P_2^\mu$ 
respectively.  Each tau has decay products (one or more neutrinos) which are unobservable.  We cannot measure the momenta of these 
decay products, but we denote hypothesised values for them as $Q_1^\mu$ and $Q_2^\mu$.
We have not been able to obtain a simple algebraic form for $m_{\tau\tau}^{\mathrm{Higgs-bound}}$; 
instead we evaluate it via a computer algorithm.
The relevant code is available from the authors on request. 
}:
\begin{equation}\label{eq:bound_def}
m_{\tau\tau}^{\mathrm{Higgs-bound}} = {\min_{\left\{Q_1^\mu,Q_2^\mu\mid\aleph\right\}}\sqrt{H^\mu H_\mu}}
\end{equation}
where
\begin{equation}
H^\mu = P_1^\mu+Q_1^\mu+P_2^\mu+Q_2^\mu 
\end{equation}
is the four momentum sum of the measured visible $P^\mu_{1,2}$ and hypothesised invisible $Q^\mu_{1,2}$ momenta of the daughters of the two taus, and where $Q_1^\mu$ and $Q_2^\mu$ are subject to constraints $\aleph$ comprising: four internal mass 
constraints
\begin{eqnarray}\label{eq:constraints_begin}
Q_1^\mu Q_{1\mu} &=& 0,  \\
Q_2^\mu Q_{2\mu} &=& 0, \\
(Q_1^\mu+P_1^\mu)(Q_{1\mu}+P_{1\mu}) &=& m_\tau^2,  \label{eq:constraints_tau1}\\
(Q_2^\mu+P_2^\mu)(Q_{2\mu}+P_{2\mu}) &=& m_\tau^2,  \label{eq:constraints_tau2}
\end{eqnarray}
and one constraint on the missing transverse momentum two-vector
\begin{equation}
\vec q_{\rm 1T}+\vec q_{\rm 2T} = \mptvec. \label{eq:constraints_end}
\end{equation}
Finally we note that it may be shown that there exists at least one pair of momenta $Q_1^\mu$ and $Q_2^\mu$ satisfying all the constraints if and only if 
\begin{equation}M_{\rm T2} (P_1,P_2,\mptvec) < m_\tau\label{eq:cond}\end{equation} 
where $M_{\rm T2}$ is the stransverse mass \cite{Lester:1999tx,Barr:2003rg,Cheng:2008hk}.  
Accordingly, it is necessary to impose a pre-selection \eqref{eq:cond} on events before $\newvar$ can be computed.

Note the difference between the design of $\newvar$ and the design of
another maximal lower-bound kinematic variable, $\mttrue$,
\cite{Barr:2009mx} which was proposed for Higgs mass measurement in
$h\rightarrow W W \rightarrow l \nu l \nu$ events.\footnote{Note that
$\mttrue$ is exactly the same as the older ``cluster transverse mass''
of \cite{Barger:1987re,Gross:2009wia} which was proposed for
the same purpose.  It is regrettable that a new notation for an
existing quantity was introduced in \cite{Barr:2009mx}, whose authors
were sadly not aware of \cite{Barger:1987re,Gross:2009wia} at time of
publication. Without prejudice to earlier work, we nonetheless retain
the $\mttrue$ notation to keep a consistent notation with the papers
to which this work is most closely tied:
\cite{Barr:2009mx,Barr:2011ux}.}  The key difference between the
$\tau\tau$ and the $WW$ topologies (other than the obvious fact that
$m_\tau\ll m_W$) is that the $W$'s need not be near their mass shells,
particularly when $m_h<2 m_W$.  Consequently $\mttrue$ does not
enforce the intermediate $W$ mass-shell constraint.\footnote{Note that
one can define a variable that applies only {\em one} internal
$W$-mass constraint.  Such a variable may be better than $\mttrue$ at
measuring $m_h$ when $m_h<2 m_W$, however it is unlikely to be better
for Higgs {\em discovery} since there is no resonant background of the
form $Z\rightarrow W W \rightarrow l \nu l \nu$ that needs to be
suppressed. For an example of a variable that is {\em not} constructed
as a mass-bound variable in the context of the $h\rightarrow W W
\rightarrow l \nu l \nu$ channel, see $m_H^{\mathrm{maos}}$ defined in
\cite{Choi:2010dw}.}

\mysubsection{Simulations}

\begin{figure*}
 \begin{center}
  \subfigure[]{\includegraphics[width=0.35\linewidth]{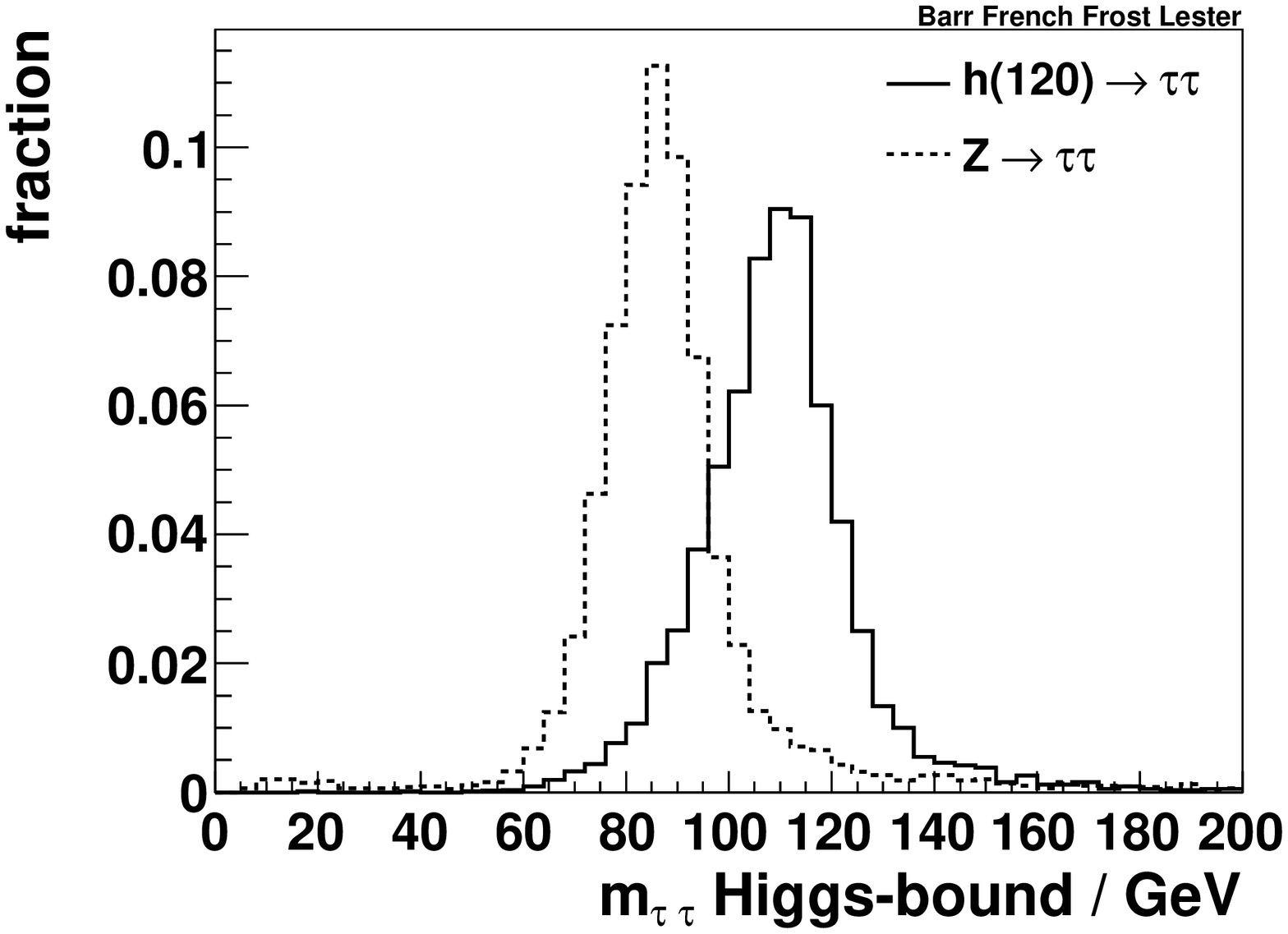}\label{fig:ex:bound}}
  \subfigure[]{\includegraphics[width=0.35\linewidth]{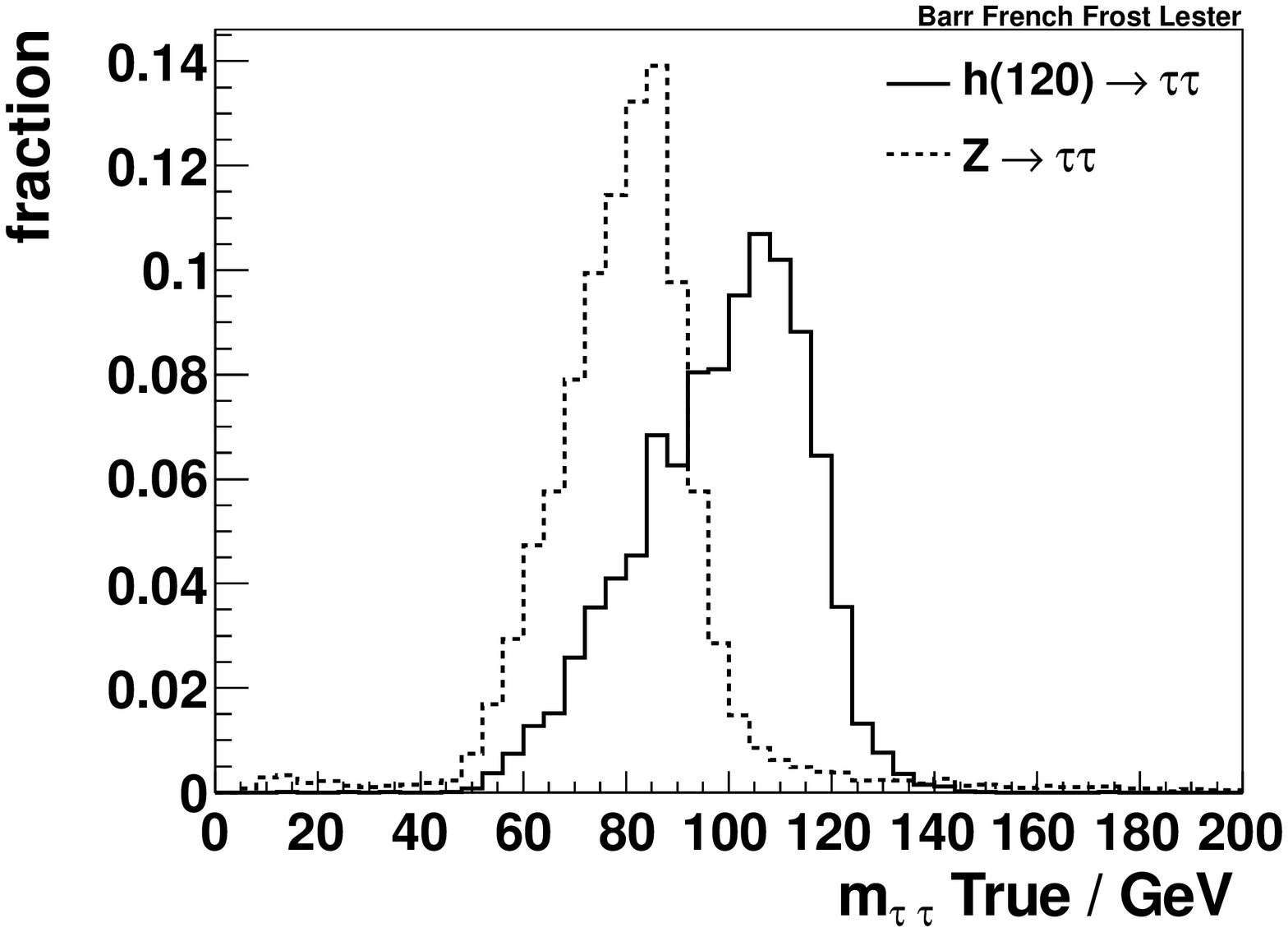}\label{fig:ex:true}}
\vskip 2mm
  \subfigure[]{\includegraphics[width=0.35\linewidth]{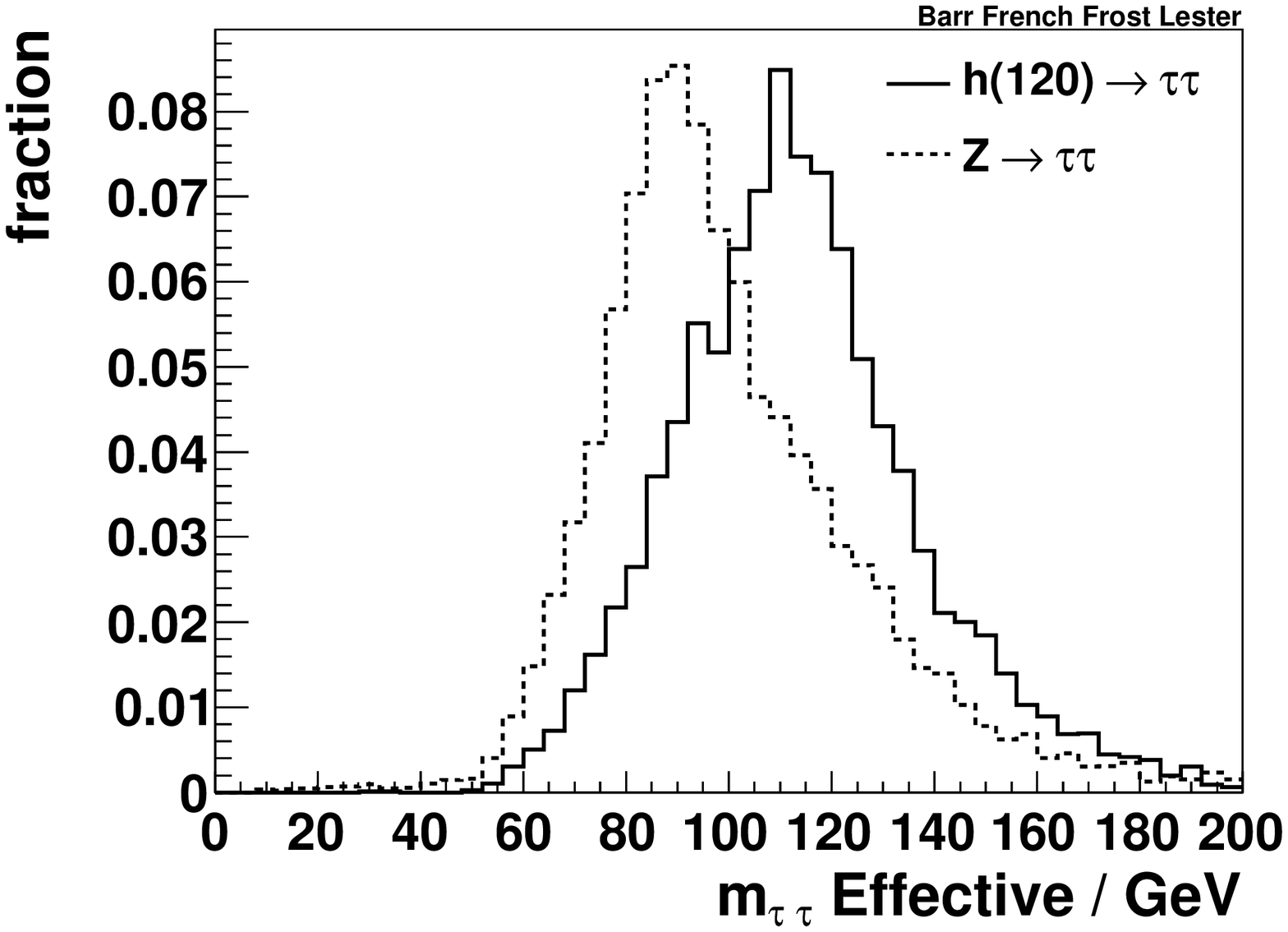}\label{fig:ex:eff}}
  \subfigure[]{\includegraphics[width=0.35\linewidth]{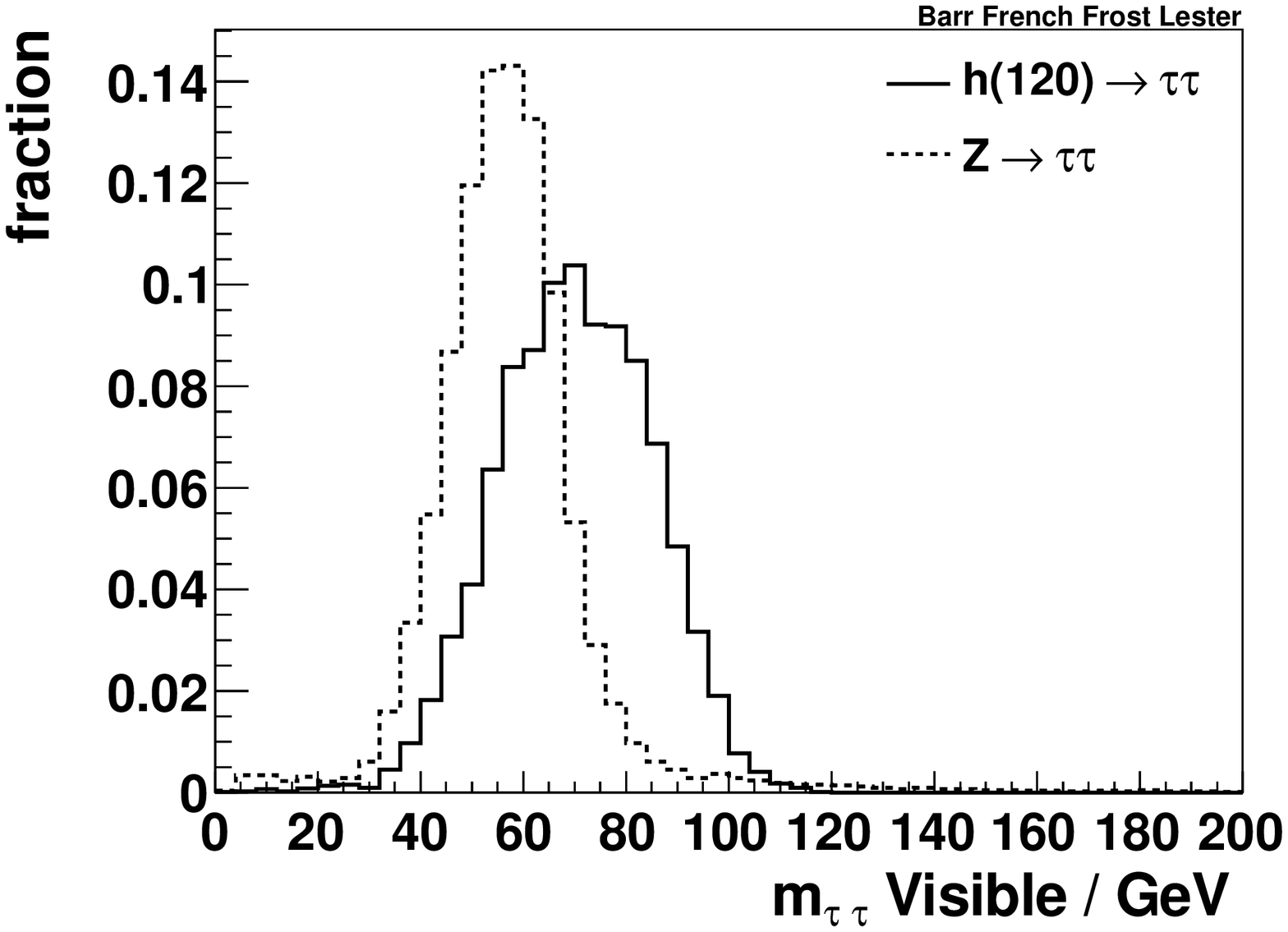}\label{fig:ex:vis}}
\caption{
The distributions of the various discriminating variables, at detector level and after the selection cuts for the simulated 120~GeV $h\rightarrow \tau\tau$ signal (solid) and $Z\rightarrow\tau\tau$ background (dashed).
The new variable
$\newvar$ \subref{fig:ex:bound} is shown, as is the transverse mass
$\mttrue$ \subref{fig:ex:true}
and two other variables currently employed by the LHC collaborations, 
$\mtautaueff$ \subref{fig:ex:eff} and 
$\mtautauvis$ \subref{fig:ex:vis}.
All signal and background histograms are presented with unit normalization to allow comparison of shape.
}
\label{fig:example_distributions}
\end{center}
\end{figure*}

\begin{figure}
 \begin{center}
  \includegraphics[width=0.99\linewidth]{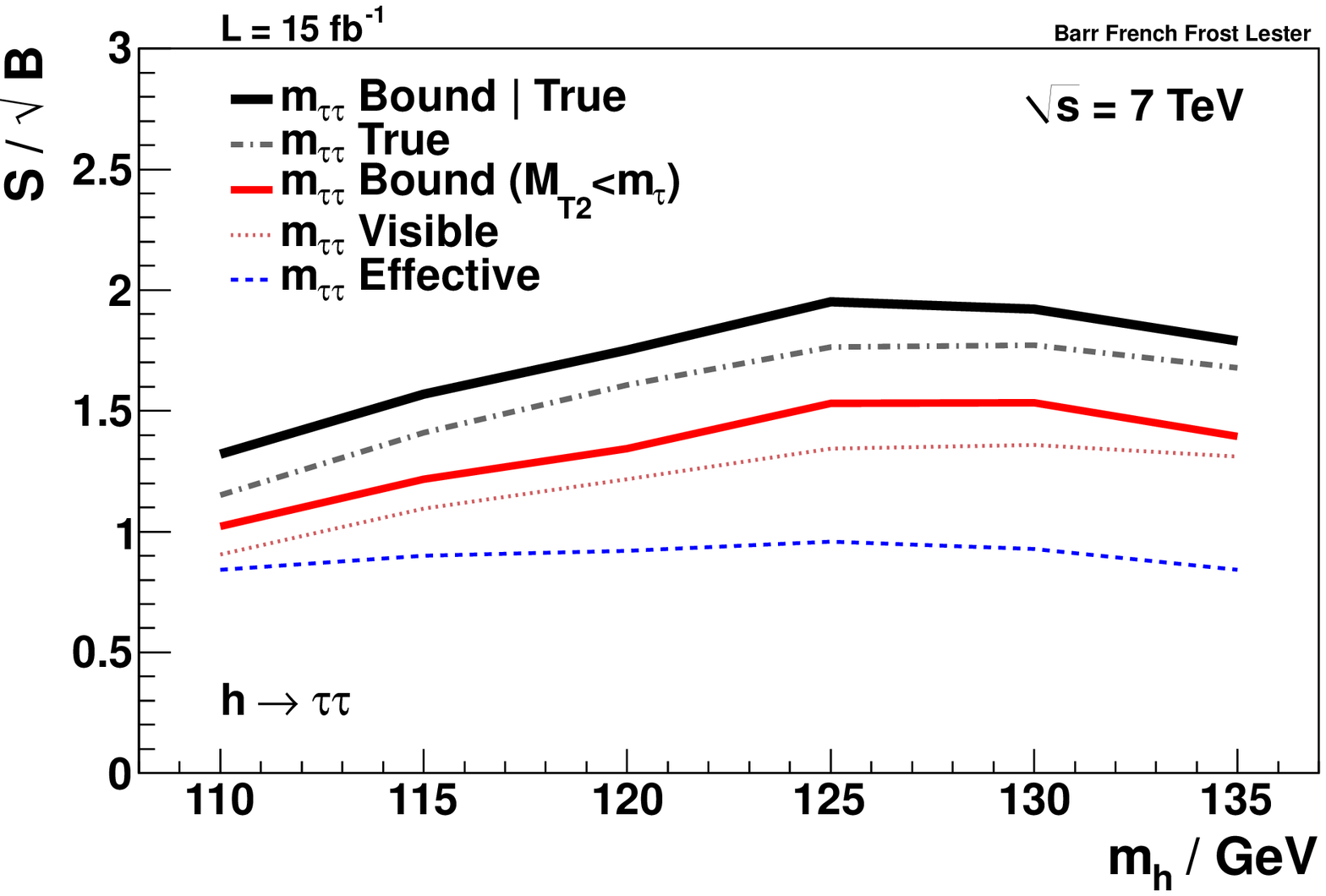}
  \caption{
Optimal discovery potential ($S/\sqrt{B}$) using each of the different discriminating variables.
The best discrimination is found using $\newvar$ for those events where it exists and $\mttrue$ for the remainder.
}
\label{fig:discovery}
\end{center}
\end{figure}

To compare the performance of $\newvar$ against other mass-scale variables, we simulate both the signal process $h \rightarrow \tau\tau$ and the dominant background $Z^0\rightarrow \tau\tau$ using the {\tt HERWIG}~6.505~\cite{Corcella:2002jc,Marchesini:1991ch} Monte Carlo generator, with LHC beam conditions ($\sqrt{s}=7~\rm{TeV}$). 

The generated tau leptons can decay either leptonically (e.g. $\tau^-\rightarrow e^-\bar{\nu}_e\nu_\tau$) or hadronically (e.g. $\tau^-\rightarrow X \nu_\tau$, where $X$ consists of hadrons or their subsequent decay products). The momenta of the visible daughters from the tau lepton decays -- electrons, muons, hadrons and photons -- ought to be well-measured by the LHC experiments. By contrast the contribution of the neutrinos to $\mptvec$ must be inferred from the negative sum of the momenta of {\em all} observed particles and so can vary considerably from its ideal value. In our simulations the missing transverse momentum is reconstructed as 
\[
\mptvec = - \sum_j\vec{p}_{\rm T,j}^{\rm\,jet}  -\sum_i \vec{p}_{\rm T,i},
\] 
where the first sum runs over all reconstructed jets, and the second runs over any stable particles within fiducial pseudorapidity 
($|\eta|<5$) and momentum ($\pt>0.5\GeV$) that are not clustered into jets.

The jets used to calculate $\vec{\slashed{p}}_{\rm T}$ are reconstructed using the fastjet~\cite{Cacciari:2005hq} implementation of the anti-$k_{\rm T}$ algorithm \cite{Cacciari:2008gp}, using the $E$ combination scheme, with distance parameter $R=0.6$ and minimum jet $\pt$ of $15\GeV$. Their energies are smeared by a Gaussian probability density function of width $$\sigma(E)/E_j = \left(0.6 \GeV^\half/\sqrt{E_j}\right)\, \oplus\, 0.03$$ where $E_j$ is the unsmeared jet energy. This resolution is typical of one of the general-purpose LHC detectors~\cite{Aad:2009wy,cmsphystdr}.

In this illustrative example, all combinations of hadronic and leptonic tau decays are treated on the same footing. We select 
events that contain two taus with pseudorapidity satisfying $|\eta|<2.5$, 
the typical angular acceptance of the tracking detector. We require that the 
visible decay products (whether electrons, muons or tau-jets) have $\pt>20$~GeV, and that $\mpt>20$~GeV.

As noted earlier, when plotting $\newvar$ we additionally require \eqref{eq:cond} to
ensure the existence of a minimisation domain in \eqref{eq:bound_def}. 
We note that in the narrow-width limit, 
well-measured tau pair events should satisfy \eqref{eq:cond} by construction. 
Therefore $\newvar$ is guarenteed to exist in the idealized case.
Detector resolution effects can be expected to lead to some events failing to satisfy \eqref{eq:cond}.
In our simulations, the consistency requirement \eqref{eq:cond} rejects about 30\% of the remaining
events from both the signal and the $Z^0\rightarrow\tau\tau$ background sample.

Example distributions for $\newvar$ (and for a number of other existing kinematical variables, described later) can be found at Monte Carlo truth level in Figure~\ref{fig:truth} and after basic detector simulation in Figure~\ref{fig:example_distributions}.
We recall that a perfect, hermetic, detector would guarentee that $\newvar \leq m_h$ for the signal and similarly $\newvar \leq m_Z$ for the $Z\rightarrow \tau\tau$ background. 
When detector resolution effects are added (Figure~\ref{fig:example_distributions}) we observe a small tail of events creeping above the ideal bound due to the smearing of the missing transverse momentum. 
Nevertheless, the signal and background $\newvar$ distributions fall off rapidly above $m_h$ and $m_Z$ respectively. 
We note that because $\newvar$ makes use of the full set of kinematic constraints (\ref{eq:constraints_begin}--\ref{eq:constraints_end}),
the space over the hypothesized momenta can be chosen during the minimisation is reduced. 
This leads to a large fraction of the events lying close to the the upper kinematic end-point.
The resulting distributions are then sharply peaked, and show good separation between the Higgs Boson signal 
and the dominant Standard Model background process.

Figure \ref{fig:example_distributions} also shows unit-normalized distributions of three
other mass-sensitive variables.
The transverse mass $\mttrue$ shown in Figure~\ref{fig:ex:true} is defined 
in a similar manner to $\newvar$, but
lacks the $\tau$ mass-shell constraints \eqref{eq:constraints_tau1}--\eqref{eq:constraints_tau2}.
Figure~\ref{fig:ex:eff} shows the distribution of the so-called ``tau-tau effective mass'', 
defined by \cite{kuhn-blois}
\begin{equation*}
\left(\mtautaueff\right)^2 = (P_1^\mu + P_2^\mu + R^\mu ) (P_{1\mu} + P_{2\mu} + R_\mu )
\end{equation*}
where $R^\mu = (\mpt,\mptvec,0)$ is a massless four-vector constructed from the missing transverse momentum.
This variable has been used as a discriminant for fully leptonic ($e^\pm,\mu^\mp$) tau events.
In our simulations it generates broad distributions with rather 
poor separation between signal and background.
Figure~\ref{fig:ex:vis} shows the distributions for the invariant mass of the visible decay products
\begin{equation*}
\left(\mtautauvis\right)^2 = (P_1^\mu + P_2^\mu)( P_{1\mu} + P_{2\mu} ),
\end{equation*}
a discriminant that has been used in observation of $Z\rightarrow \tau\tau$~\cite{ATLAS-CONF-2011-010} and searches for MSSM Higgs bosons~\cite{ATLAS-CONF-2011-024,cms_higgs_tau_tau}.
Again, the distributions are broader, 
and the separation between signal and background less pronounced, than for $\newvar$.

We quantify the discrimination power of each of the four variables 
as a function of $m_h$ as follows.
For each variable 
$f \in \{\mtautaueff$, $\mtautauvis$, $\newvar$, $\mttrue\}$
we plot the distribution, after detector simulation and the cuts described above, 
of the signal and of the dominant $Z\rightarrow \tau\tau$ background.
We then determine the value of the additional cut $f^{\rm min}$ that maximises
$S/\sqrt{B}$ subject to $S>10$, where $S$ and $B$ are the numbers
of signal and background events, respectively, 
that would be expected with $f>f^{\rm min}$
for 15\,fb$^{-1}$ of integrated luminosity.

The optimal values of $S/\sqrt{B}$ obtained for the four different projections
are shown as a function of $m_h$ in Figure~\ref{fig:discovery}.
The line marked $\newvar$ is the result obtained if one simply discards
the $\sim$30\% of events that fail to satisfy \eqref{eq:cond}.
and hence for which $\newvar$ is not defined.

Both $\mttrue$ and $\newvar$ perform better than the two currently-employed alternatives.
We might expect $\newvar$, which has a sharper distribution than $\mttrue$, 
to have a larger significance.
However, because it is only possible to define $\newvar$
for events that satisfy \eqref{eq:cond}, its stand-alone significance is reduced.
What one really wants is to combine the desirable features of the two related mass-bound variables ---
the sharper distribution of $\newvar$, and the guarenteed existance,
even after smearing, of $\mttrue$.
One way of doing so is by forming a `best effort' variable, defined to be
$\newvar$ if it exists, and $\mttrue$ otherwise.
The resulting distribution is found to have the largest value of $S/\sqrt{B}$ 
for all values of $m_h$ simulated.

The peformance of the Bound-or-True combined variable has been compared to yet another alternative, one obtained from an event-by-event maximization of a likelihood over all allowable neutrino momenta~\cite{Elagin:2010aw}. 
The distribution obtained using the method of Ref.~\cite{Elagin:2010aw} results in an almost equally high discovery potential, but unlike our proposal it must be tuned to the particular kinematic cuts empolyed, and it is a factor of $>$1000 more computationally expensive to compute.

\mysubsection{Conclusions}

We have advocated use of $\newvar$ to separate $h\rightarrow\tau\tau$ events from their most significant irreducible Standard Model background $Z\rightarrow \tau\tau$. 
We observe that, for events in which a solution to the full set of kinematic constraints exists,
the superior discriminatory power of $\newvar$ arises by construction: it is as {\em the} maximal lower bound on an important scale (in this case the invariant mass of the tau pair), which is smaller than $m_Z$ for the background, but is often $\sim m_h$ for the signal. 

For events with no $\newvar$ solutions,
the usual transverse mass $\mttrue$ can be substituted in the same role, 
so that one makes good use of {\em all} events.
Choosing $\newvar$ when available and $\mttrue$ otherwise offers superior discriminatory power than either 
variable alone,
and in all cases a better discovery potential 
than the variables currently being used in LHC searches.

Although we have focused our attention here on the decay $h\rightarrow \tau\tau$, it is worth remembering that the same variable can be used on any topology with a similar kinematic structure having a resonance decaying to a pair of intermediate on-shell particles under the substitution $m_\tau \rightarrow m_{\mathrm{intermediate}}$.


\begin{acknowledgments}
We are grateful to Chris Boddy, Ben Gripaios, Claire Gwenlan and Trevor Vickey for assistance and advice.
This work was supported by the Science and Technology Research Council of the United Kingdom, by the Royal Society, by Merton College, Oxford and by Peterhouse, Cambridge.
\end{acknowledgments}
\bibliography{HTT}
\end{document}